\documentclass[12pt]{article}  %fuer beidseitige Version
\usepackage[a4paper,textwidth=490.0pt,textheight=703.1pt]{geometry}
\usepackage{slashed}
\usepackage{graphicx}
\usepackage{epsfig}
\usepackage{amsmath}
\usepackage{amssymb}
\usepackage{ulem}
\usepackage{mdwlist}
\usepackage{color}                                                  
\usepackage{float}
\usepackage[rflt]{floatflt}
\usepackage{slashed}          
\usepackage{cite}
\usepackage{mathtools}
\usepackage{dsfont}
\usepackage{subfig}
\usepackage{array}

\usepackage[english]{babel}
\usepackage[latin1]{inputenc}
\usepackage[T1]{fontenc}
\usepackage{ae}

\usepackage{url}

\usepackage{amsmath, amsthm, amssymb}

\newtheorem{thm}{Theorem}[section]

\usepackage{array}

\usepackage[english]{babel}
\usepackage[latin1]{inputenc}
\usepackage[T1]{fontenc}
\usepackage{ae}

\newtheorem{definition}[thm]{Definition}

\newcommand{\ep}{\varepsilon}

\usepackage{rotating}

\usepackage{graphicx}
\newcounter{mmacnt}
\def\restartmma{\setcounter{mmacnt}{0}}
\restartmma \catcode`|=\active
\def|#1|{\mathrm{#1}}
\catcode`|=12
\newenvironment{mma}{
 \par\smallskip
 \catcode`|=\active
 \parskip=0pt\parindent=0pt % locally
 \small
 \def\In##1\\{%
   \def\linebreak{\hfill\break\null\qquad}%
   \refstepcounter{mmacnt}
   \hangindent=2.5em\hangafter=0
   \leavevmode
   \llap{\tiny\sffamily In[\arabic{mmacnt}]:=\kern.5em}%
   \mathversion{bold}\footnotesize$\displaystyle##1$\normalsize
   \mathversion{normal}\par
 }%
 \def\Print##1\\{%
   \def\linebreak{\hfill\break}%
   \hangindent=2.5em\hangafter=0
   \leavevmode ##1\par}%
 \def\Out##1\\{%
   \def\linebreak{$\hfill\break\null\hfill$}%
   \kern\abovedisplayskip\par
   \hangindent=2.5em\hangafter=0
   \leavevmode
   \llap{\tiny\sffamily Out[\arabic{mmacnt}]=\kern.5em}
   \footnotesize$\displaystyle##1$\normalsize\hfill\null\par
   \kern\belowdisplayskip
 }%
 \def\Warning##1##2\\{%
   \def\linebreak{\hfill\break}%
   \hangindent=2.5em\hangafter=0
   \leavevmode
   {\scriptsize##1 : ##2}\par}%
}{%
 \par\smallskip
}

\allowdisplaybreaks[4]

\usepackage{color}

\newenvironment{fshaded}{%
\MakeFramed {\FrameRestore}
}%
{\endMakeFramed}

\usepackage{tikz}
\usetikzlibrary{matrix}

\allowdisplaybreaks[4]

\begin{document}
\setlength{\baselineskip}{0.515cm}
\sloppy
\thispagestyle{empty}
\begin{flushleft}
DO--TH  23/15 \hfill  % {\tt arXiv:24XX.XXXXX [hep-ph]}
\\
DESY 24--027 \hfill March 2024\\
CERN-TH-2024-030 \\
RISC Report series 24--02\\
ZU-TH 13/24 
\end{flushleft}

\setcounter{table}{0}

\begin{center}

\vspace*{1cm}
{\Large\bf \boldmath The non-first-order-factorizable contributions to the three-loop 
single-mass operator matrix elements $A_{Qg}^{(3)}$ and $\Delta A_{Qg}^{(3)}$}

\vspace{\fill}
\large
J.~Ablinger$^{a,b}$, 
A.~Behring$^{c}$,
J.~Bl\"umlein$^{d,e}$,
A.~De Freitas$^{d,a}$, \\
A.~von Manteuffel$^{f}$,
C.~Schneider$^a$,
and
K.~Sch\"onwald$^{g}$

\vspace*{1cm}
{\small
{\it $^a$~Johannes Kepler University, 
Research Institute for Symbolic Computation (RISC), \newline
Altenberger Stra\ss{}e 69,
                          A-4040, Linz, Austria}

\vspace*{2mm}
{\it $^b$~Johann Radon Institute for Computational and Applied Mathematics
(RICAM), \newline
Austrian Academy of Sciences, 
Altenberger Stra\ss{}e 69, A-4040 Linz Austria}

\vspace*{2mm}
{\it $^c$~Theoretical Physics Department, CERN, 1211 Geneva 23, Switzerland}

\vspace*{2mm}
{\it $^d$~Deutsches Elektronen-Synchrotron DESY, Platanenallee 6, 15738 Zeuthen, Germany}

\vspace*{2mm}
{\it $^e$ Institut f\"ur Theoretische Physik III, IV, TU Dortmund, Otto-Hahn Stra\ss{}e 4, \newline 44227 
Dortmund,
Germany}

\vspace*{2mm}
{\it $^f$~Institut f\"ur Theoretische Physik, Universit\"at Regensburg,
93040 Regensburg, Germany}

\vspace*{2mm}
{\it $^g$~Physik-Institut, Universit\"at Z\"urich, 
Winterthurerstrasse 190, CH-8057 Z\"urich, Switzerland}}
\\

\end{center}
\normalsize
\vspace{\fill}
\begin{abstract}
\noindent 
The non-first-order-factorizable contributions\footnote{The terms 'first-order-factorizable contributions'
and 'non-first-order-factorizable contributions' have been introduced and discussed in Refs.~\cite{Behring:2023rlq,
Ablinger:2023ahe}. They describe the factorization behaviour of the difference- or differential
equations for a subset of master integrals of a given problem.}  to the unpolarized and polarized massive operator 
matrix elements to three-loop order, $A_{Qg}^{(3)}$ and $\Delta A_{Qg}^{(3)}$, are calculated in the single-mass 
case. For the $_2F_1$-related master integrals of the problem, we use a semi-analytic method based
on series expansions and utilize the first-order differential equations for the master integrals which
does not need a special basis of the master integrals. Due to the singularity structure of this basis a part 
of the integrals has to be computed to $O(\ep^5)$ in the dimensional parameter. The solutions have to be 
matched at a series of thresholds and pseudo-thresholds in the region of the Bjorken variable $x \in ]0,\infty[$ 
using highly precise series expansions to obtain the imaginary part of the physical amplitude for $x \in ]0,1]$ 
at a high relative accuracy. We compare the present results both with previous analytic results, the results 
for fixed Mellin moments, and a prediction in the small-$x$ region. We also derive expansions in 
the region of small and large values of $x$. With this paper, all three-loop single-mass unpolarized and 
polarized operator matrix elements are calculated.
\end{abstract}

\vspace*{\fill}
\numberwithin{equation}{section}

\newpage
%%%%%%%%%%%%%%%%%%%%%%%%%%%%%%%%%%%%%%%%%%%%%%%%%%%%%%%%%%%%%%%%%%%%%%%%%%%%%%%%%%%%%%%%%%%%%%%%%%%
\section{Introduction}
\label{sec:1}
%%%%%%%%%%%%%%%%%%%%%%%%%%%%%%%%%%%%%%%%%%%%%%%%%%%%%%%%%%%%%%%%%%%%%%%%%%%%%%%%%%%%%%%%%%%%%%%%%%%

\vspace*{1mm}
\noindent
The heavy-flavor contributions both to the unpolarized and polarized deep-inelastic structure functions
form an essential part of these quantities. Their scaling violations are different from those of the massless
contributions. Since the experimental precision reached the 1\% level in the unpolarized case with HERA
\cite{Blumlein:1989pd}, which will also be the case for polarized deep-inelastic scattering at EIC 
\cite{Boer:2011fh} and the proposed LHeC \cite{LHeCStudyGroup:2012zhm}, the three-loop heavy-flavor 
corrections are needed in the QCD analysis of these data.

In a previous paper \cite{Ablinger:2023ahe}, we have calculated the first-order factorizable contributions to 
the constant parts of the  unrenormalized three-loop massive operator matrix elements (OMEs) $A_{Qg}^{(3)}$ 
and $\Delta A_{Qg}^{(3)}$   ($a_{Qg}^{(3)}$ and $\Delta a_{Qg}^{(3)}$), which are based on 1009 of a
total of 1233 Feynman diagrams. For all contributions at least 1000 non-vanishing Mellin 
moments are known. 
Furthermore, 15 of the 25 color-$\zeta$ contributions of the OMEs have been calculated by using the
method of arbitrarily high Mellin moments \cite{Blumlein:2017dxp}. 
Here $\zeta_n, n \in \mathbb{N}, n \geq 2$ denote the values of Riemann's $\zeta$-function at integer values 
of $n$. Their associated recurrences were 
computed by using the guessing method \cite{GUESS,Blumlein:2009tj} and solved using the package {\tt Sigma}
\cite{SIG1,SIG2} in all first-order-factorizing cases.

The major new aspect of the present calculation concerns the contribution of higher transcendental letters
in the iterated integrals forming the master integrals given by $_2F_1$-solutions \cite{Ablinger:2017bjx}
\footnote{They are related to complete elliptic integrals and modular forms, 
cf.~Refs.~\cite{SABRY,Broadhurst:1987ei,Broadhurst:1993mw,Bloch:2013tra,Adams:2015ydq,
Remiddi:2016gno,Adams:2017ejb,Broedel:2019hyg} and the surveys in 
Ref.~\cite{Blumlein:2019tmi}.} 
of Heun differential equations \cite{HEUN1,HEUN2}. The corresponding basic integrals have been 
computed to $O(\ep^0)$ in Ref.~\cite{Behring:2023rlq}. Here $\ep =  D-4$ denotes the dimensional 
parameter. The other master integrals can be obtained by 
iterating Kummer-Poincar\'e \cite{KUMMER1,KUMMER2,KUMMER3,POINCARE,LANDAN,CHEN,GONCHAROV,Borwein:1999js,
Moch:2001zr,Ablinger:2013cf} and square-root valued letters \cite{Ablinger:2014bra,Ablinger:2023ahe} on the former 
solutions, 
by which all master integrals can be obtained. Future work will be devoted to this method of iterating 
non-iterative integrals, including higher transcendental letters.

In the present paper, we are following a different avenue. Here we {do not compute the analytic results for 
the previously mentioned
OMEs, but we obtain highly precise semi-analytic results}.
{To derive these results, we}
start with the {coupled} first-order differential equation system of master integrals,
{which is} obtained {from} the integration by parts (IBP) reduction 
\cite{Studerus:2009ye,vonManteuffel:2012np}.
The solution of these equations is performed in $t$-space \cite{Ablinger:2012qm,Ablinger:2014yaa}
{via series expansions around various points which are numerically matched in
overlapping regions of convergence}. Here $t$ denotes the resummation parameter.
{The initial conditions are given at $t=0$ as the
Mellin moments to the required order in $\ep$. In our set of master integrals one has to
calculate up to $O(\ep^5)$ in individual cases. The initial values have been computed already
before for determining the corresponding recurrences in Mellin-$N$ space, 
cf.~\cite{Blumlein:2017wxd,Ablinger:2023ahe}.}
After analytic continuation from $t$ to
$x$-space, cf.~\cite{Behring:2023rlq}, {one obtains the final expression}.
The analytic continuation has to pass
a series of pseudo-thresholds and thresholds from $x \rightarrow \infty$ {($t=0$)} to 
$x = 0$ {($t \rightarrow \infty$)} and matching
conditions have to be evaluated. The present approach uses large mantissa rational matching in this
process. In this way, we finally obtain the constant parts of the unrenormalized massive OMEs
$A_{Qg}^{(3)}$ and $\Delta A_{Qg}^{(3)}$,  $a_{Qg}^{(3)}$ and $\Delta a_{Qg}^{(3)}$. This formalism
is only applied to the part of the amplitude which is affected by $_2F_1$-related
letters.

The paper is organized as follows. In Section~\ref{sec:2} we describe the basic computation method.
In Section~\ref{sec:3} we compute $a_{Qg}^{(3)}$ and $\Delta a_{Qg}^{(3)}$ in $x$-space, compare to
previous partial results in the literature, and present numerical results. The results for small and 
large values of Bjorken $x$ are presented in Section~\ref{sec:4} and Section~\ref{sec:5} contains 
the conclusions. 
%%%%%%%%%%%%%%%%%%%%%%%%%%%%%%%%%%%%%%%%%%%%%%%%%%%%%%%%%%%%%%%%%%%%%%%%%%%%%%%%%%%%%%%%%%%%%%%%%%%
\section{The main steps of the calculation}
\label{sec:2}
%%%%%%%%%%%%%%%%%%%%%%%%%%%%%%%%%%%%%%%%%%%%%%%%%%%%%%%%%%%%%%%%%%%%%%%%%%%%%%%%%%%%%%%%%%%%%%%%%%%

\vspace*{1mm}
\noindent
The calculation of the contributing Feynman diagrams from their generation to the reduction to the master
integrals has been described in Ref.~\cite{Ablinger:2023ahe}. Here we use the packages {\tt QGRAF, Form, Color} and 
{\tt Reduze~2} \cite{Nogueira:1991ex,Vermaseren:2000nd,Tentyukov:2007mu,vanRitbergen:1998pn,Studerus:2009ye,
vonManteuffel:2012np}, 
and apply the Feynman rules given in Refs.~\cite{YND,Bierenbaum:2009mv}. In the polarized 
case we compute the OME in the Larin scheme \cite{Larin:1993tq}. The OMEs are calculated using the method
described in  Ref.~\cite{Behring:2023rlq}.
This means the operators defined for discrete integer values of $N$
are resummed into a generating function which depends on the continuous real variable $t$.
The master integrals are computed in this
variable by solving linear systems of coupled differential equations, 
see also Refs.~\cite{Ablinger:2018zwz,
Maier:2017ypu,Fael:2021kyg,Fael:2022miw,Ablinger:2023ahe}. The initial values are provided by the Mellin 
moments \cite{Blumlein:2017wxd}, which are the expansion coefficients at $t=0$. The system of differential 
equations is solved at a series of thresholds and pseudo-thresholds in the region $t \in [0,\infty[$.
The variables $t$ and the Bjorken variable $x$ are related by
%-------------------------------------------------------------------------------------------------------
\begin{eqnarray}
x = \frac{1}{t}.
\end{eqnarray}
%-------------------------------------------------------------------------------------------------------
The set of thresholds and pseudo-thresholds in the differential equations of all master integrals, resp.
the necessary expansion points, given the convergence radius of the  respective local series, are
%-------------------------------------------------------------------------------------------------------
\begin{eqnarray}
x \in \Biggl\{0, \frac{1}{32}, \frac{1}{16}, \frac{1}{8}, \frac{1}{6}, \frac{1}{4}, \frac{1}{2},
\frac{2}{3}, \frac{3}{4}, \frac{5}{6}, \frac{8}{9}, 1\Biggr\}
\end{eqnarray}
%-------------------------------------------------------------------------------------------------------
for $x \in [0,1]$ in the present problem. Imaginary parts of the amplitude develop only at the transition 
point $x = 1$, and for no other points in the regions $x \in ]0,1[$ and $x \in ]1,\infty[$.
The expansion points for $x > 1$ were
%-------------------------------------------------------------------------------------------------------
\begin{eqnarray}
x \in \Biggl\{\frac{8}{7}, \frac{4}{3}, 2, 4, \infty\Biggr\}. 
\end{eqnarray}
%-------------------------------------------------------------------------------------------------------
The analysis starts at $x = \infty$ using  the previously computed moments as initial values.
The expansion order in $\ep$ depends on the individual master integral.
In the present case one needs to expand up to $O(\ep^5)$ for some integrals.
One performs series expansions around the \mbox{(pseudo-)}thresholds by inserting a suitable
ansatz into the differential equation.
Comparing coefficients in $\ep$, the expansion parameter $t-t_0$ (and possible powers of logarithms) one
obtains a large system of equations for the symbolic expansions.
This system of linear equations is solved with {\tt FireFly} \cite{Klappert:2019emp,Klappert:2020aqs}
using modular methods in terms of a small number of boundary constants.
These are determined by matching two neighboring expansions in the middle with 250 digits 
accuracy.
In order to solve this linear system, we rationalize the arising floating point numbers, which allows 
{for a stable solution}.
Except for the expansions around $x=0$ and $x=1$, we compute 100 expansion coefficients, while at the
latter points, which contain in addition powers of the logarithms $\ln(x)$ and $\ln(1-x)$, respectively,
50 expansion terms are used.

The initial values at $x \rightarrow \infty$ are real, as are also the coefficients of the linear differential
equation systems. For the expansion points for {$x > 1$}, the series expansions 
are real-valued {Taylor} series
in $x$ and their contribution to the massive OMEs vanish, see~\cite{Behring:2023rlq}. {The 
analytic
continuation at $x=1$ implies logarithmic-modulated series containing powers of
$\ln^k(1-x)$ and thus an imaginary part after analytic continuation.
The imaginary part is proportional to the $x$-space representation of the massive OMEs, 
see~Ref.~\cite{Behring:2023rlq}.}
%%%%%%%%%%%%%%%%%%%%%%%%%%%%%%%%%%%%%%%%%%%%%%%%%%%%%%%%%%%%%%%%%%%%%%%%%%%%%%%%%%%%%%%%%%%%%%%%%%%
\section{\boldmath $a_{Qg}^{(3)}(x)$ and $\Delta a_{Qg}^{(3)}(x)$}
\label{sec:3}
%%%%%%%%%%%%%%%%%%%%%%%%%%%%%%%%%%%%%%%%%%%%%%%%%%%%%%%%%%%%%%%%%%%%%%%%%%%%%%%%%%%%%%%%%%%%%%%%%%%

\vspace*{1mm}
\noindent
In Ref.~\cite{Ablinger:2023ahe}, we computed all color-$\zeta$ contributions which can be obtained by
solving difference equations that factorize into first-order factors. 
Furthermore, we calculated all remaining irreducible
Feynman diagrams with contributions from master integrals, whose associated differential equations 
factorize into first-order factors.
The remaining 224 Feynman diagrams 
are related to $_2F_1$-solutions \cite{Ablinger:2017bjx,Behring:2023rlq} and are calculated in the present
paper by solving the first order differential equations obtained from the IBP-relations directly
in a highly precise numerical approach, adding the previous results to the complete solution.

On top of the irreducible Feynman diagrams mentioned before, there are also reducible Feynman diagrams and 
ghost contributions to the amplitudes $A_{Qg}^{(3)}(x)$ and 
$\Delta A_{Qg}^{(3)}(x)$, contributing to the final result, which we would like to characterize 
briefly. In the unpolarized case, in Mellin $N$-space, they are spanned by the harmonic and generalized 
harmonic sums~\cite{Vermaseren:1998uu,Blumlein:1998if,Ablinger:2013cf}
%---------------------------------------------------------------------------------------------------------------------------$
\begin{eqnarray}
&& \Biggl\{
S_{-4},S_{-3},S_{-2},S_1,S_2,S_3,S_4,S_{-3,1},S_{-2,1},S_{-2,2},
S_{3,1},S_{2,1},S_{2,1,1},S_{-2,1,1}, 2^{N},
S_1\Biggl(\Biggl\{\frac{1}{2}\Biggr\}\Biggr),
\nonumber\\ &&
S_1(\{2\}),
S_{1,3}\Biggl(\Biggl\{\frac{1}{2},2\Biggr\}\Biggr),
S_{2,1}(\{1,2\}),
  S_{2,1}(\{2,1\}),
    S_{1,1,1}(\{2,1,1\}),
      S_{1,1,2}\Biggl(\Biggl\{\frac{1}{2},2,1\Biggr\}\Biggr),
\nonumber\\ &&
      S_{1,2,1}\Biggl(\Biggl\{\frac{1}{2},2,1\Biggr\}\Biggr),
        S_{1,1,1,1}\Biggl(\Biggl\{\frac{1}{2},2,1,1\Biggr\}\Biggr)\Biggr\}
\end{eqnarray}
%---------------------------------------------------------------------------------------------------------------------------$
with rational and $2^{N}$-prefactors. In $x$-space, these terms convert to harmonic polylogarithms 
\cite{Remiddi:1999ew}
at argument $x$ or $1 - 2x$, as in Ref.~\cite{Ablinger:2014nga}. The generalized sums stem from the ghost 
contributions, which are absent in the polarized case.

Expanding their contribution around $x = 0$  terms of order
%---------------------------------------------------------------------------------------------------------------------------$
\begin{eqnarray}
\frac{\ln^4(x)}{x},~~
\frac{\ln^3(x)}{x},~~
\frac{\ln^2(x)}{x}
\end{eqnarray}
%---------------------------------------------------------------------------------------------------------------------------$
with
%---------------------------------------------------------------------------------------------------------------------------$
\begin{eqnarray}
\textcolor{blue}{C_A^2 T_F} 
\frac{1}{x}\Biggl[\frac{1}{54} \ln^4(x) + \frac{1}{18} \ln^3(x) + 
\left(
\frac{61}{36} + \frac{1}{3} \zeta_2\right)\ln^2(x)\Biggr],
\end{eqnarray}
%---------------------------------------------------------------------------------------------------------------------------$
are present, which are not expected in the complete result. Here the color factors are 
$\textcolor{blue}{C_A} = N_c, \textcolor{blue}{C_F} = (N_c^2-1)/(2 N_c), \textcolor{blue}{T_F} 
= 1/2$ for $SU(N_c)$ and $N_c = 3$ for 
Quantum Chromodynamics (QCD).  Indeed, the calculation shows
that these terms are canceled at a relative accuracy of 
%---------------------------------------------------------------------------------------------------------------------------$
\begin{eqnarray}
\left\{-2.7134 \cdot 10^{-17}, -1.1975 \cdot 10^{-13},
-1.4327 \cdot 10^{-15}\right\}
\end{eqnarray}
%---------------------------------------------------------------------------------------------------------------------------$
in the complete result numerically. Contributions of this kind do not emerge in the polarized case.

Our present results  can be tested also in various other ways. Next we compare the result in $x$-space with 
the moments computed in Ref.~\cite{Bierenbaum:2009mv} by a totally different method,
using {\tt MATAD} \cite{Steinhauser:2000ry}. For the moments $N = 2,4,6,8,10$ we obtain agreement up to
relative accuracies of
%---------------------------------------------------------------------------------------------------------------------------$
\begin{eqnarray}
\Biggl\{-4.3039 \cdot 10^{-8},
1.0758 \cdot 10^{-9},
6.9438 \cdot 10^{-10},
-4.3401 \cdot 10^{-11},
-1.4872 \cdot 10^{-10}
\Biggr\}
\end{eqnarray}
%---------------------------------------------------------------------------------------------------------------------------$
in the unpolarized case. Since the first moment of $\Delta a_{Qg}^{(3)}$ turns out to be zero, we compare
here the relative deviation of the moments $N = 3,5,7,9,11$, for which we obtain
%---------------------------------------------------------------------------------------------------------------------------$
\begin{eqnarray}
\Biggl\{
-8.9221 \cdot 10^{-10},
 9.6270 \cdot 10^{-10},
-2.4977 \cdot 10^{-10},
-1.7849 \cdot 10^{-10},
 3.1817 \cdot 10^{-11}
\Biggr\}.
\end{eqnarray}
%---------------------------------------------------------------------------------------------------------------------------$
%---------------------------------------------------------------------------------------------------------------------------$
\begin{figure}[H]
\centering
\includegraphics[width=0.49\textwidth]{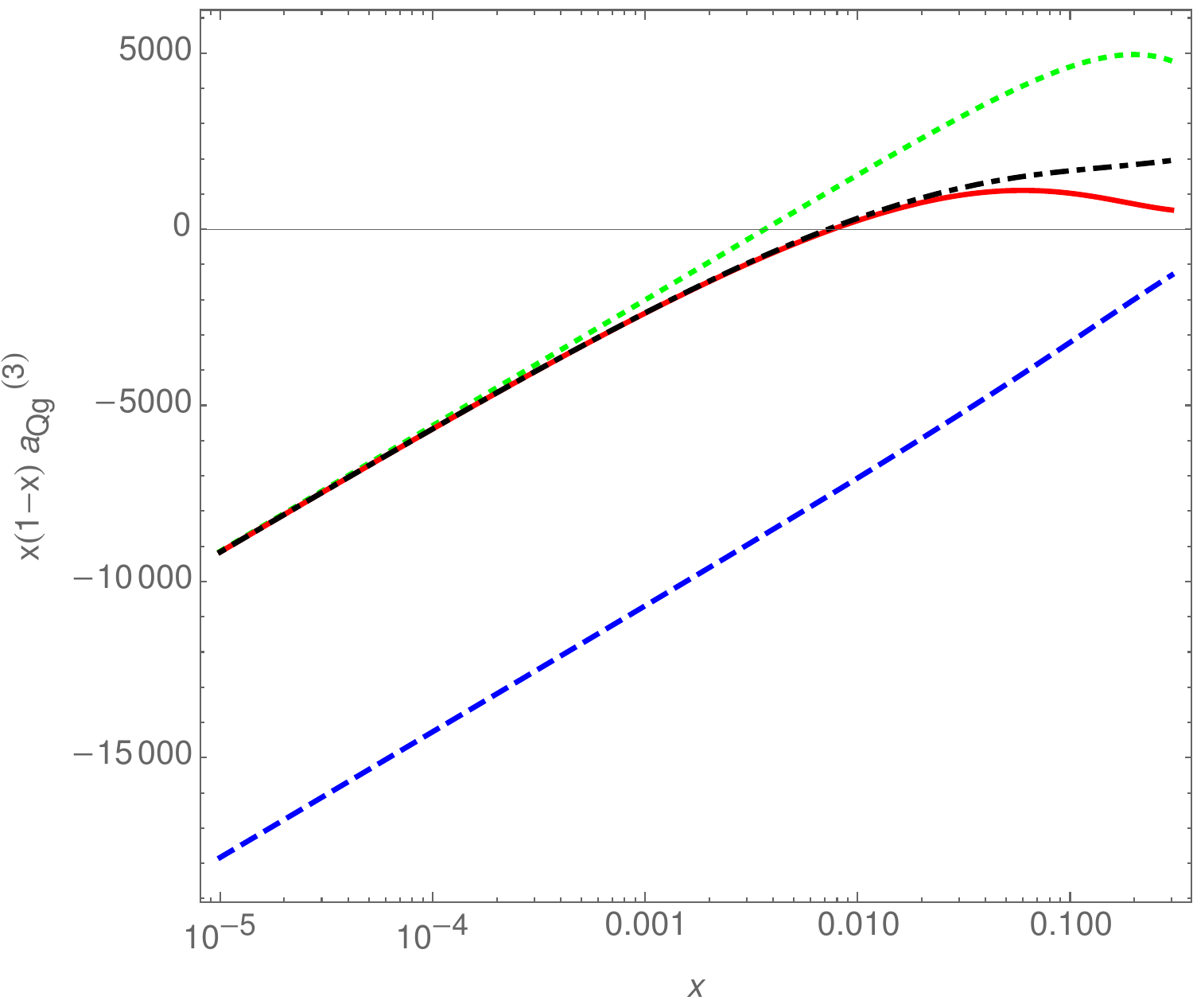}
\includegraphics[width=0.49\textwidth]{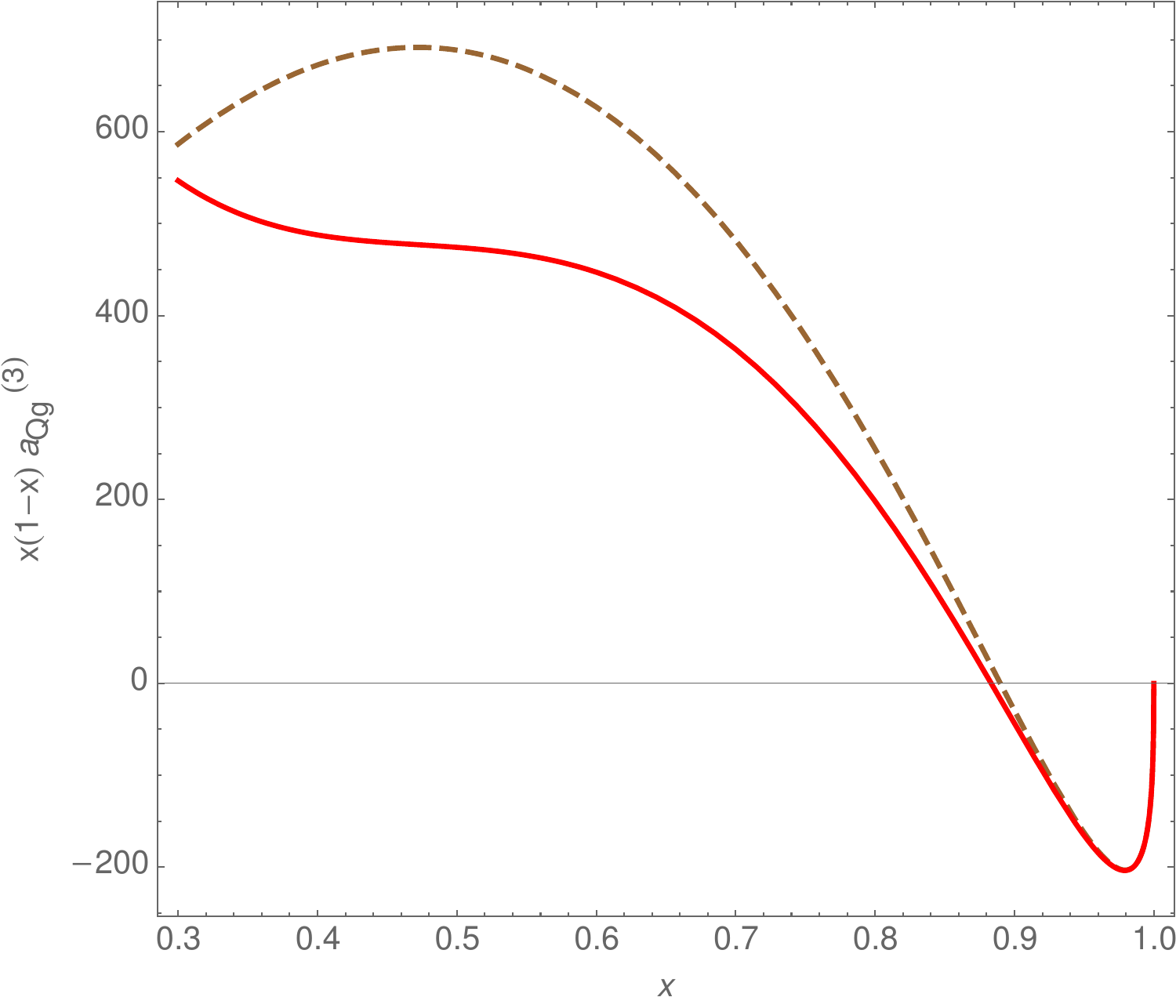}
\caption{\sf 
$a_{Qg}^{(3)}(x)$  as a function of $x$, rescaled by the factor $x(1-x)$.
Left panel: smaller $x$ region.
Full line (red): $a_{Qg}^{(3)}(x)$;
dashed line (blue): leading small-$x$ term $\propto \ln(x)/x$ 
\cite{Catani:1990eg};
dotted line (green): $\ln(x)/x$ and $1/x$ term;
dash-dotted line (black): all small-$x$ terms, including also 
$\ln^k(x)$,~\mbox{$k \in \{1,\ldots,5\}$}.
Right panel: larger $x$ region. 
Full line (red): $a_{Qg}^{(3)}(x)$; dashed line (brown): leading large-$x$ terms up to the  
terms $\propto (1-x)$, covering the logarithmic contributions of $O(\ln^k(1-x))$,~\mbox{$k \in \{1,\ldots, 4\}$}.
}
\label{fig:1}
\end{figure}
%---------------------------------------------------------------------------------------------------------------------------
A further test of accuracy consists in the comparison of the present differential equation 
method with the analytic results obtained by $N$-space techniques for the \textcolor{blue}{$N_F$} terms 
in $a_{Qg}^{(3)}(x)$ and $\Delta a_{Qg}^{(3)}(x)$ before, where
\textcolor{blue}{$N_F$} denotes the number of massless flavors.  For
%---------------------------------------------------------------------------------------------------------------------------$
\begin{eqnarray}
(\Delta) r(x) = \frac{(\Delta) a_{Qg}^{(3), N_F,\rm deq}(x)}{(\Delta) a_{Qg}^{(3), N_F,\rm ex}(x)} - 1
\end{eqnarray}
%---------------------------------------------------------------------------------------------------------------------------$
we obtain 
%---------------------------------------------------------------------------------------------------------------------------$
\begin{eqnarray}
x &\rightarrow& \Biggl\{
\frac{1}{100}, 
\frac{1}{10}, 
\frac{3}{10}, 
\frac{5}{10}, 
\frac{7}{10}, 
\frac{9}{10}, 
\frac{99}{100}\Biggr\},
\\
r(x) &\rightarrow& \Biggl\{
-1.66 \cdot 10^{-17},
-1.18 \cdot 10^{-16},
-4.97 \cdot 10^{-16},
-4.01 \cdot 10^{-16},
-1.88 \cdot 10^{-15},
\nonumber\\ &&
-4.42 \cdot 10^{-17},
8.56 \cdot 10^{-18} \Biggr\}.
\\
\Delta r(x) &\rightarrow& \Biggl\{
-2.91 \cdot 10^{-17},
 9.09 \cdot 10^{-16},
-1.71 \cdot 10^{-15},
-1.38 \cdot 10^{-15},
-1.88 \cdot 10^{-15},
\nonumber\\ &&
7.89 \cdot 10^{-16},
-1.06 \cdot 10^{-15}
\Biggr\}.
\end{eqnarray}
%---------------------------------------------------------------------------------------------------------------------------$

%---------------------------------------------------------------------------------------------------------------------------$
\begin{figure}[H]
\centering
\includegraphics[width=0.49\textwidth]{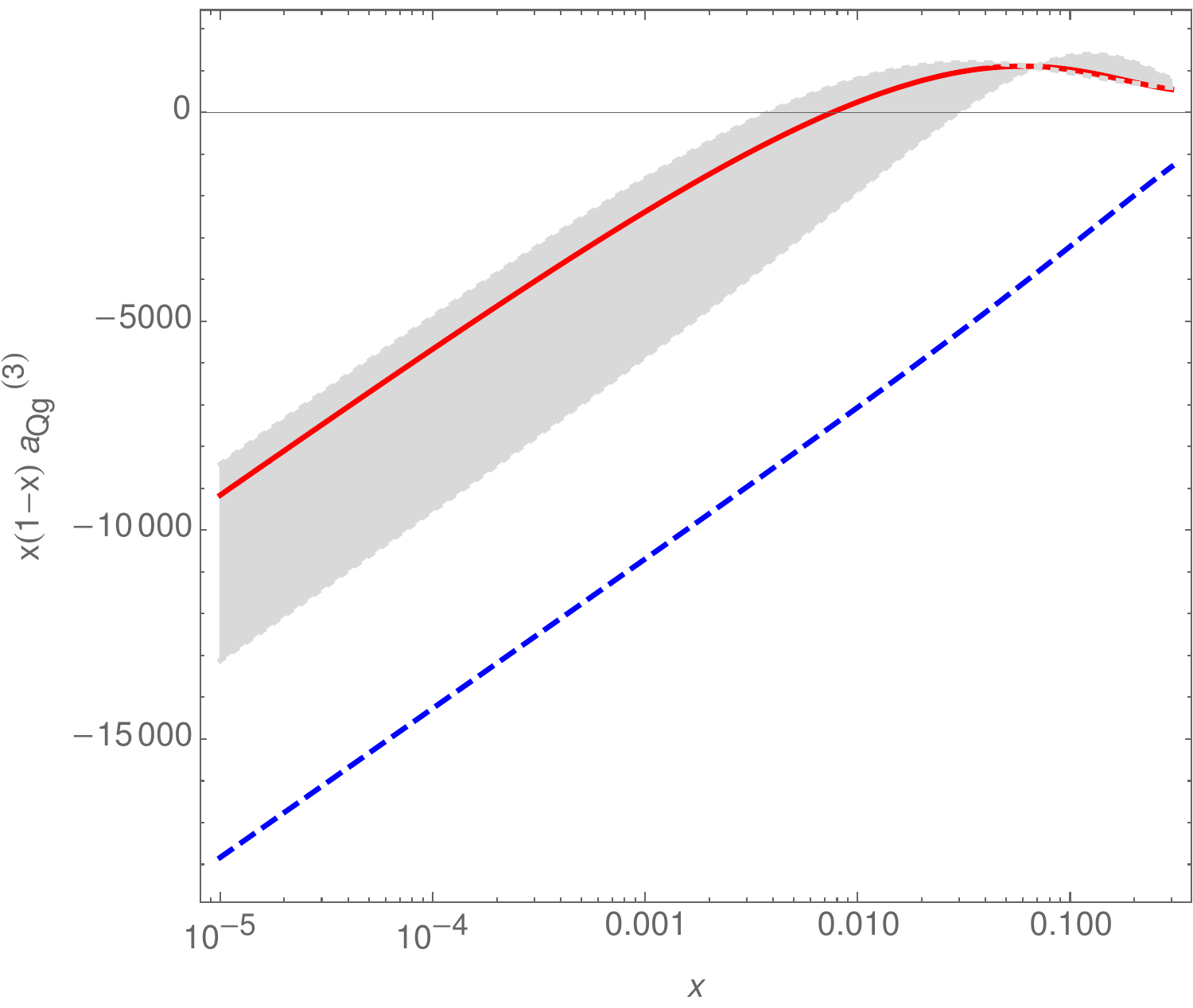}
\includegraphics[width=0.49\textwidth]{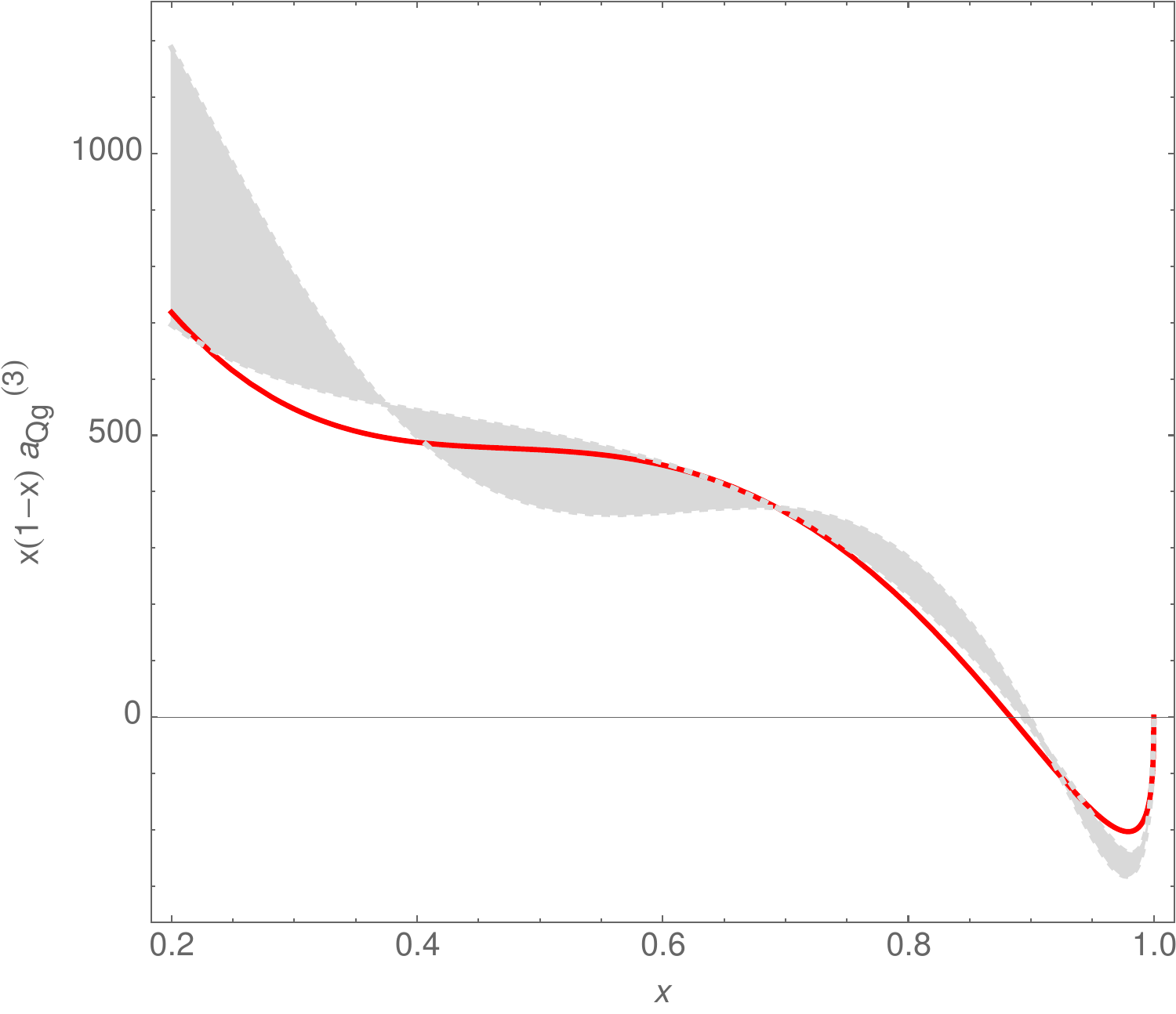}
\caption{\sf 
$a_{Qg}^{(3)}(x)$  as a 
function of $x$, rescaled by the factor $x(1-x)$.
Left panel: smaller $x$ region.
Full line (red): $a_{Qg}^{(3)}(x)$;
dashed line (blue): leading small-$x$ term $\propto \ln(x)/x$ 
\cite{Catani:1990eg};
gray region: estimates of \cite{Kawamura:2012cr}.
Right panel: larger $x$ region. 
Full line (red): $a_{Qg}^{(3)}(x)$; 
gray region: estimates of \cite{Kawamura:2012cr}.
}
\label{fig:2}
\end{figure}
%---------------------------------------------------------------------------------------------------------------------------
%---------------------------------------------------------------------------------------------------------------------------$
\begin{figure}[H]
\centering
\includegraphics[width=0.49\textwidth]{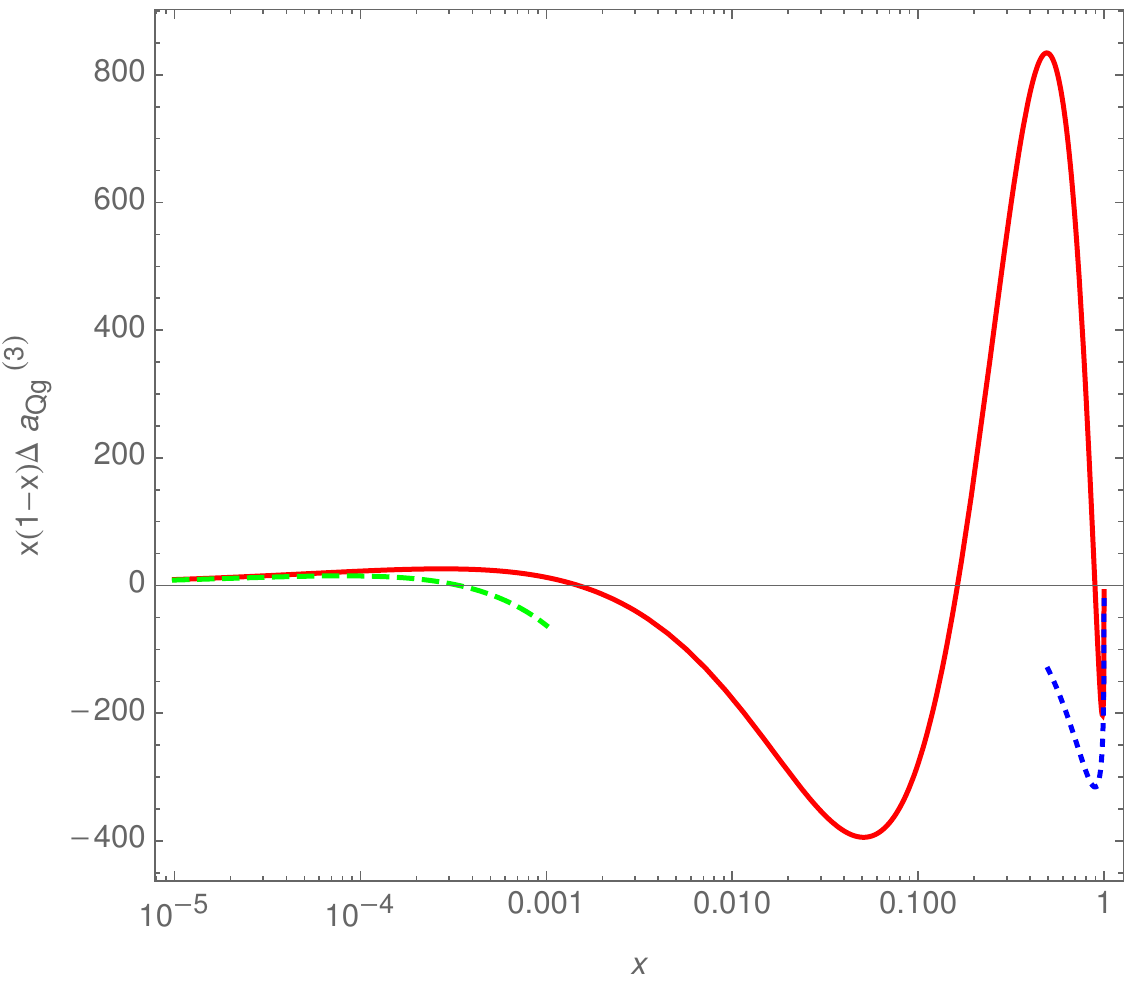}
\includegraphics[width=0.49\textwidth]{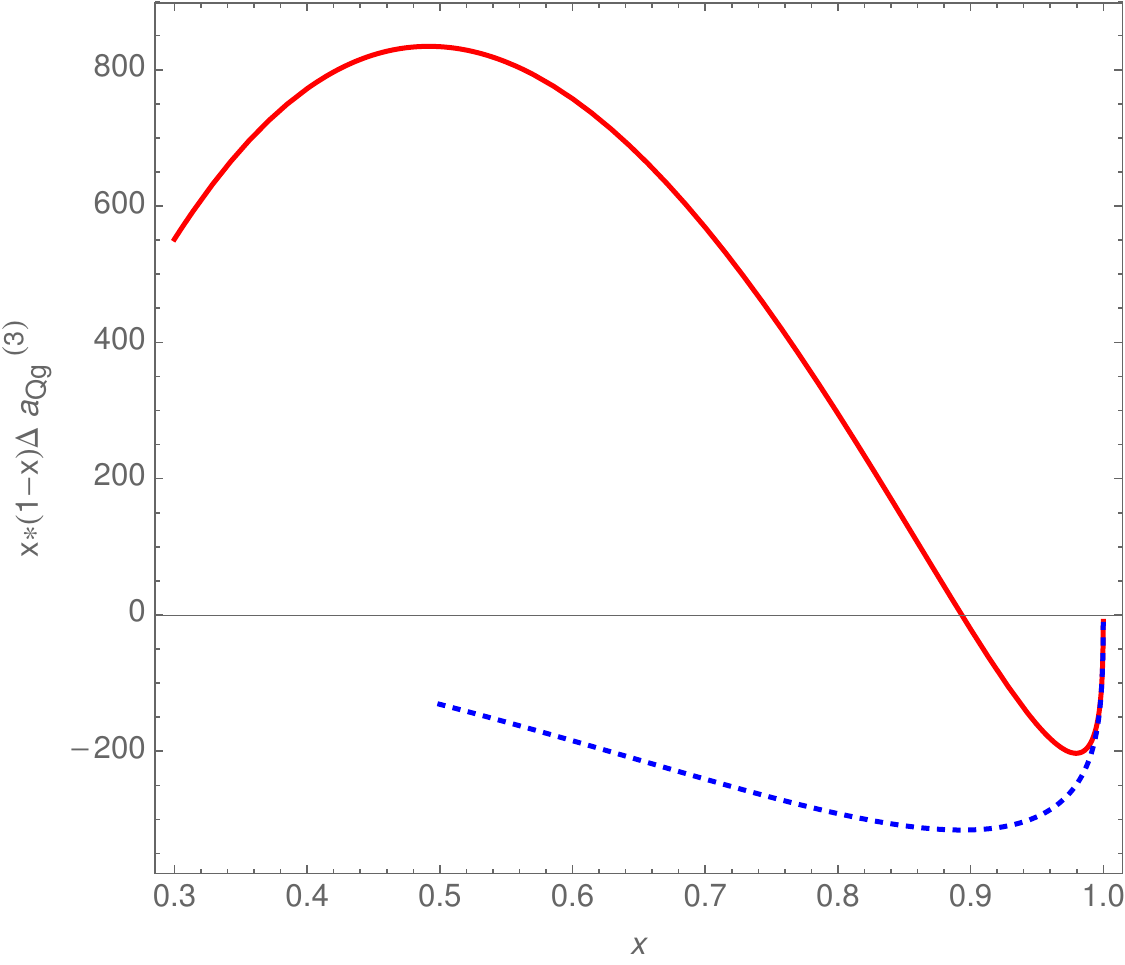}
\caption{\sf 
$\Delta a_{Qg}^{(3)}(x)$  as a 
function of $x$, rescaled by the factor $x(1-x)$.
Left panel: full line (red): $\Delta a_{Qg}^{(3)}(x)$;
dashed line (green): the small-$x$ terms $\ln^k(x)$, \mbox{$k \in \{1,\ldots,5\}$};
dotted line (blue): the large-$x$ terms $\ln^l(1-x)$, \mbox{$l \in \{1,\ldots,4\}$}.
Right panel: larger $x$ region. 
Full line (red): $\Delta a_{Qg}^{(3)}(x)$; dotted line (blue): 
the large-$x$ terms $\ln^l(1-x)$, \mbox{$l \in \{1,\ldots,4\}$}.
}
\label{fig:3}
\end{figure}
%---------------------------------------------------------------------------------------------------------------------------

\noindent
For smaller values of $x$, the deviations are even smaller.
A comparable accuracy has been obtained for the pole terms of the unrenormalized 
amplitudes $A_{Qg}^{(3)}(x)$ and $\Delta A_{Qg}^{(3)}(x)$, which are also known in analytic 
form.

Now we turn to the results of $a_{Qg}^{(3)}(x)$ and $\Delta a_{Qg}^{(3)}(x)$.
In Figures~\ref{fig:1} and \ref{fig:2} we illustrate the analytic result for $a_{Qg}^{(3)}(x)$ for 
QCD and 
by setting $\textcolor{blue}{N_F} = 3$ in the smaller and larger $x$ regions.
Here we add different lines for the small-$x$ and large-$x$ 
contributions to show their validity.
\noindent
The so-called leading small-$x$ result \cite{Catani:1990eg} turns out 
not to describe
the physical quantity $a_{Qg}^{(3)}(x)$ quantitatively; see, however, the discussion in 
Section~\ref{sec:4}. This is, as in all known other cases, see 
e.g.~\cite{Blumlein:1995jp,Blumlein:1996hb,Blumlein:1997em,Blumlein:1999ev,Ablinger:2014nga},
due to sub-leading terms which cancel the leading behaviour. 

Here the inclusion of the $1/x$ 
term, not predicted by small-$x$ methods, leads to a description up 
to $x \sim 10^{-4}$.
To describe the region to $x \sim 2 \cdot 10^{-2}$ one needs also all contributing 
$\ln^k(x)$-terms, with $k \in \{1,\ldots,5\}$. In the large-$x$ region no expansion terms have been 
predicted.
Here one obtains a description down to $x \sim 0.9$ by considering all $\ln^k(1-x)$ terms for $k 
\in \{1,\ldots,4\}$, including the $(1-x)^0$ and $(1-x)$ contributions.

In Ref.~\cite{Kawamura:2012cr} estimates on the size of the charm quark contributions in 
$F_2(x,Q^2)$ were made based on five moments for $A_{Qg}^{(3)}$ and six moments for 
$A_{Qq}^{(3),\rm PS}$ calculated in Ref.~\cite{Bierenbaum:2009mv}, the two-loop contributions 
of Refs.~\cite{Buza:1995ie,Bierenbaum:2007qe}, and the $N_F$-terms from our calculation in 
Ref.~\cite{Ablinger:2010ty}. Furthermore, the small-$x$ behaviour 
from \cite{Catani:1990eg} for  $A_{Qg}^{(3)}$ and a corresponding color-rescaled leading small-$x$ term for 
$A_{Qq}^{(3),\rm PS}$ were assumed. The latter has only later  been proven in Ref.~\cite{Ablinger:2014nga} by 
calculating $A_{Qq}^{(3),\rm PS}$ in complete form analytically. In \cite{Kawamura:2012cr} 
the three other contributing OMEs $A_{qg,Q}^{(3)}$, $A_{qq,Q}^{(3), \rm NS}$, $A_{qq,Q}^{(3), 
\rm PS}$, as well as the two-mass corrections, were not taken into account.
In Figure~\ref{fig:2} we illustrate the former estimate on $a_{Qg}^{(3)}(x)$ in the region 
of smaller and larger values of $x$ (gray band) and compare it to the exact result (red 
lines), which lie close to the upper end of the former estimate in the region of small values of $x$.

Let us now turn to the polarized case. The quantity $\Delta a_{Qg}^{(3)}(x)$ is shown in Figure~\ref{fig:3},
where we also indicate the small- and large-$x$ terms, cf.~Section~\ref{sec:4}. We note that the 
latter approximations match the exact result only in the extreme regions.
$\Delta a_{Qg}^{(3)}(x)$ shows an oscillatory behaviour as also known for the polarized structure 
functions $g_{1,2}(x,Q^2)$. 
One reason for this behaviour is that the first moment of $\Delta a_{Qg}^{(3)}(x)$ vanishes. Also
at the first and second order in the strong coupling constant the first moment vanishes,
cf.~Ref.~\cite{Bierenbaum:2022biv}.

%%%%%%%%%%%%%%%%%%%%%%%%%%%%%%%%%%%%%%%%%%%%%%%%%%%%%%%%%%%%%%%%%%%%%%%%%%%%%%%%%%%%%%%%%%%%%%%%%%%
\section{The small- \boldmath and large-$x$ limits}
\label{sec:4}
%%%%%%%%%%%%%%%%%%%%%%%%%%%%%%%%%%%%%%%%%%%%%%%%%%%%%%%%%%%%%%%%%%%%%%%%%%%%%%%%%%%%%%%%%%%%%%%%%%%

\vspace*{1mm}
\noindent
In this section, we discuss in more detail the small- and large-$x$ limits which have already been 
illustrated in Section~\ref{sec:3}. The leading small-$x$ contribution to the unpolarized 
quantity $a_{Qg}^{(3)}(x)$ has been predicted
in Ref.~\cite{Catani:1990eg}, within a leading order calculation based on $k_\perp$-factorization.
The result is given by  
%-------------------------------------------------------------------------------------------------
\begin{eqnarray}
\label{eq:SX1}
a_{Qg}^{(3), x \rightarrow 0}(x) &=& \frac{64}{243} \textcolor{blue}{C_A^2 T_F} \left[
1312 + 135 \zeta_2 - 189 \zeta_3\right] \frac{\ln(x)}{x}.
\end{eqnarray}
%-------------------------------------------------------------------------------------------------
As we saw in Section~\ref{sec:3}, this result is interesting for the theoretical comparison to the 
corresponding
term in the complete calculation, but cannot be used for phenomenology due to the destructive sub-leading 
corrections. We obtained the term $\propto \zeta_2$ in Ref.~\cite{Ablinger:2023ahe} since it results from
first order factorizing contributions only. From the small-$x$ expansion of the present 
result, we obtain an agreement on the purely rational and $\zeta_3$ term of Eq.~(\ref{eq:SX1})
at a relative accuracy of 
%-------------------------------------------------------------------------------------------------
\begin{eqnarray}
\label{eq:SX2}
\{ -8.0143 \cdot 10^{-16} \}.
\end{eqnarray}
%-------------------------------------------------------------------------------------------------
This is the first independent recalculation of the result of Ref.~\cite{Catani:1990eg} using
a different method, and it also establishes the rescaling to the corresponding analytic result in the 
pure-singlet case \cite{Ablinger:2014nga}.

Since we knew the coefficient of the $\zeta_2$-term, cf.~\cite{Ablinger:2023ahe}, in (\ref{eq:SX1}), the question is 
whether
integer relations allow to determine the other two terms in (\ref{eq:SX1}) at the level of 15 known digits. At 
least there could be a conditional answer in assuming a certain rational prime factor pattern, starting out with 
$2, 3, 5, 7$ to reasonably small powers watching out for matches by using the LLL algorithm \cite{LLL} {\tt 
lindep} in the package {\tt Pari} \cite{PARI2}. David Broadhurst has been so kind to do this 
for us. He obtained
%-------------------------------------------------------------------------------------------------
\begin{eqnarray}
\textcolor{blue}{C_A^2 T_F}\left[\frac{2^{11} \cdot \textcolor{red}{41}}{3^5}-\frac{2^6 \cdot 7}{3^2} 
\zeta_3\right]
\end{eqnarray}
%-------------------------------------------------------------------------------------------------
very quickly, by which (\ref{eq:SX1}) can be considered to be confirmed using methods of experimental
mathematics.\footnote{So, the answer to the ultimate leading small-$x$ question of inclusive heavy-flavor 
physics 
is actually
\textcolor{red}{41}. For differing answers to similar questions, see \cite{DAD}.}

The small-$x$ terms for $a_{Qg}^{(3)}$ are given numerically by\footnote{Here we present 10 digits for
brevity, although our results are more accurate.}
\begin{eqnarray}
\label{eq:SX3}
a_{Qg}^{(3), x \rightarrow 0} &\simeq&
    1548.891667 \frac{\ln(x)}{x} 
  + [8956.649545 
  - 88.20492033 \textcolor{blue}{N_F}] \frac{1}{x} 
\nonumber\\ &+&
 [4.844444444 - 0.4444444444 \textcolor{blue}{N_F}] \ln^5(x)
\nonumber\\ &+&
 [-21.75925926 - 2.506172840 \textcolor{blue}{N_F}] \ln^4(x) 
\nonumber\\ &+&
 [514.0912722 - 35.20953611 \textcolor{blue}{N_F}]  \ln^3(x)
\nonumber\\ 
&+& [-720.0483828 - 90.85414199 \textcolor{blue}{N_F}] \ln^2(x) 
\nonumber\\ &+&
 [10739.21741 - 468.0849296 \textcolor{blue}{N_F}] \ln(x). 
\end{eqnarray}
The alternating sign of the first two coefficients is the main reason why the leading small-$x$
contributions are not sufficient to describe the final result even at very 
small values of $x$. For a
precise description also the sub-leading terms are needed.

In the large-$x$ limit $a_{Qg}^{(3)}$ is given by 
%-------------------------------------------------------------------------------------------------
\begin{eqnarray}
\label{SX3a}
a_{Qg}^{(3), x \rightarrow 1} &\simeq&
3.703703704 \ln^5(1-x) +
[-8.20987654 + 0.4938271605 \textcolor{blue}{N_F}] \ln^4(1-x)
\nonumber\\ 
&+& [4.380199906 + 1.646090535 \textcolor{blue}{N_F}] \ln^3(1-x)
\nonumber\\ 
&+& [-332.5368214 - 0.4183246058 \textcolor{blue}{N_F}] \ln^2(1-x) 
\nonumber\\
&+& [ 737.165347 - 73.1297935 \textcolor{blue}{N_F}] \ln(1-x).
\end{eqnarray} 
%-------------------------------------------------------------------------------------------------

In the polarized case the leading small-$x$ terms are 
%-------------------------------------------------------------------------------------------------
\begin{eqnarray}
\label{eq:SX3b}
\Delta a_{Qg}^{(3), x \rightarrow 0} &\simeq&
   [-12.60493827 + 0.4444444444 \textcolor{blue}{N_F}] \ln^5(x)
\nonumber\\ 
&+&  [-145.2160494 + 7.839506173 \textcolor{blue}{N_F}] \ln^4(x)
\nonumber\\ 
&+&  [-856.9645724 + 63.82682006 \textcolor{blue}{N_F}] \ln^3(x)
\nonumber\\ 
&+&  [-852.7889255 + 298.2461398 \textcolor{blue}{N_F}] \ln^2(x)
\nonumber\\ 
&+&  [25006.51309 + 544.6633205 \textcolor{blue}{N_F}] \ln(x).
\end{eqnarray}
%-------------------------------------------------------------------------------------------------
Here the coefficients of the terms $\ln(x)/x$ and $1/x$ have been shown to be zero in \cite{Ablinger:2023ahe}. 
The $\textcolor{blue}{N_F}$ term $\propto \ln^5(x)$ has been derived in Ref.~\cite{Ablinger:2023ahe} as 
$\textcolor{blue}{C_F T_F^2 N_F}(4/3) \ln^5(x)$. The coefficients in (\ref{eq:SX3b}) are alternating,
except for the last term.

The expansion coefficients in the large-$x$ region are
%-------------------------------------------------------------------------------------------------
\begin{eqnarray}
\label{eq:SX4}
\Delta a_{Qg}^{(3), x \rightarrow 1} &\simeq&  
 3.703703704 \ln^5(1-x)
+ [-8.20987654 + 0.49382716105 \textcolor{blue}{N_F}] \ln^4(1-x)
\nonumber\\ 
&+& [4.380199906 + 1.646090535 \textcolor{blue}{N_F}] \ln^3(1-x)
\nonumber\\ 
&+& [-332.5368214 - 0.4183246058 \textcolor{blue}{N_F}] \ln^2(1-x)
\nonumber\\ 
&+& [737.165347 - 73.1297935 \textcolor{blue}{N_F}] \ln(1-x).
\end{eqnarray}
%-------------------------------------------------------------------------------------------------
The large-$x$ expansions, Eqs.~(\ref{SX3a}, \ref{eq:SX4}) are the same in the unpolarized and 
polarized case. The same behaviour has already been seen for the factorizing contributions before, 
cf.~\cite{Ablinger:2023ahe}. The results in the unpolarized and polarized cases were obtained by separate
calculations.
%%%%%%%%%%%%%%%%%%%%%%%%%%%%%%%%%%%%%%%%%%%%%%%%%%%%%%%%%%%%%%%%%%%%%%%%%%%%%%%%%%%%%%%%%%%%%%%%%%%
\section{Conclusions}
\label{sec:5}
%%%%%%%%%%%%%%%%%%%%%%%%%%%%%%%%%%%%%%%%%%%%%%%%%%%%%%%%%%%%%%%%%%%%%%%%%%%%%%%%%%%%%%%%%%%%%%%%%%%

\vspace*{1mm}
\noindent
We  have calculated the non-first-order-factorizable contributions to the three-loop massive
operator matrix elements $A_{Qg}^{(3)}$ and $\Delta A_{Qg}^{(3)}$ in the single-mass case. 
This completes the computation of 
these matrix elements and thereby of all of the three-loop single-mass unpolarized and polarized OMEs 
\cite{Ablinger:2010ty,
Blumlein:2012vq,
Behring:2014eya,
Ablinger:2014lka,
Ablinger:2014vwa,
Ablinger:2014nga,
Ablinger:2019etw,
Behring:2021asx,
Ablinger:2022wbb,
Ablinger:2023ahe}. Also the two-mass three-loop corrections \cite{Ablinger:2017err,
Ablinger:2017xml,
Ablinger:2018brx,
Ablinger:2019gpu,
Ablinger:2020snj,
Blumlein:2021xlc}, except those for $(\Delta) A_{Qg}^{(3)}$, have already been computed.
The solution of the first-order differential equation system of master integrals in different 
sub-intervals of $x \in ]0,\infty[$ at very high numerical precision and high precision matching
using the methods \cite{Ablinger:2018zwz,Behring:2023rlq} allowed us to derive the three-loop corrections
tied up to iterated non-iterative integrals containing $_2F_1$-letters in terms of local series
expansions. The latter are logarithmic-modulated with powers of $\ln(x)$ around $x=0$, and $\ln(1-x)$
around $x=1$.

We confirm the leading small-$x$ prediction for the $O(\ln(x)/x)$ term in the unpolarized case in an 
independent calculation using a different method for the first time. We compared our results
with the moments of Ref.~\cite{Bierenbaum:2009mv} and other terms, which were calculated by us using 
different
methods and found agreement. 

The present results are important for future
measurements of the  strong coupling constant $\alpha_s(M_Z^2)$ 
\cite{Bethke:2011tr,Moch:2014tta,Alekhin:2016evh,dEnterria:2022hzv}, 
the charm quark mass, $m_c$, \cite{Alekhin:2012vu},
and the parton distribution functions, see~e.g.~\cite{Accardi:2016ndt,Alekhin:2017kpj}.
All the three loop single-and two-mass corrections to deep-inelastic scattering will be released
in form of a numerical code in a forthcoming publication.

\vspace*{5mm}
\noindent
{\bf Acknowledgments}.\\
We would like to thank S.~Klein for calculating a series of Mellin moments in the polarized case 
\cite{SKLEIN} by using {\tt MATAD}, which we have used for comparison, and we thank P.~Marquard for 
discussions. We would like to thank D.~Broadhurst to have conditionally determined two missing rational coefficients 
of the unpolarized leading small-$x$ contribution to $a_{Qg}^{(3)}$ from just 15 digits.
This work was supported by the Austrian Science Fund (FWF) grants P33530 and P34501N.

%-----------------------------------------------------------------------------------------------------

%-----------------------------------------------------------------------------------------------------
\end{document}